\documentclass[aps,notitlepage,twocolumn,superscriptaddress,longbibliography,nofootinbib]{revtex4-1}
\pdfoutput=1
\usepackage[utf8]{inputenc}
\usepackage{amsmath,amsthm,amssymb,amsfonts}
\usepackage{scalerel}
\usepackage{mathtools}
\usepackage{multirow}
\usepackage[nice]{nicefrac}
\usepackage{amsmath}
\usepackage{xfrac}
\usepackage{graphicx}
\usepackage{subfigure}
\usepackage[normalem]{ulem}
\usepackage[colorlinks = true,
            linkcolor = blue,
            urlcolor  = blue,
            citecolor = blue,
            anchorcolor = blue]{hyperref}
\usepackage{mathrsfs}
\usepackage{bbold}
\usepackage[]{units}
\usepackage{bm}
\usepackage{braket}
\usepackage{color}
\usepackage{makecell}
\usepackage{enumitem}
\usepackage{upgreek}
\usepackage{blindtext}
\usepackage{graphics}
\usepackage{verbatim} 
\usepackage{algorithm}
\usepackage[noend]{algpseudocode}

\usepackage[dvipsnames]{xcolor}
\usepackage{bbm}
\usepackage[bb=boondox]{mathalfa}
\usepackage{array}
\usepackage{multirow}
\usepackage{tabularx}
\usepackage{float}
\usepackage{graphicx}
\usepackage{dcolumn}
\usepackage{bm}
\usepackage{amsmath,amsfonts,amssymb}
\usepackage{slashed}
\usepackage{braket,xcolor}
\usepackage{verbatim}
\usepackage{multirow}
\usepackage{amsfonts, mathtools,resizegather}
\usepackage[export]{adjustbox}

\def\H{\mathcal{H}}

\def\bra#1{\left\langle#1\right|}
\def\ket#1{\left|#1\right\rangle}

\def\abs#1{\left|#1\right|}
\def\kc#1{\left(#1\right)}

\def\ke#1{\left\{#1\right\}}

\def\be{\begin{equation}}       \def\ee{\end{equation}}
\def\bea{\begin{eqnarray}}      \def\eea{\end{eqnarray}}
\def\ba{\begin{array}}
	\def\ea{\end{array}}
\def\bnum{\begin{enumerate} }
	\def\enum{\end{enumerate}}

\def\=>{\Rightarrow}
\def\>{\rightarrow}

\def\eye2{Fathbb{I}}

\begin{document}

\title{Krylov space approach to Singular Value Decomposition in non-Hermitian systems}

\author{Pratik Nandy\,\,\href{https://orcid.org/0000-0001-5383-2458}
{\includegraphics[scale=0.05]{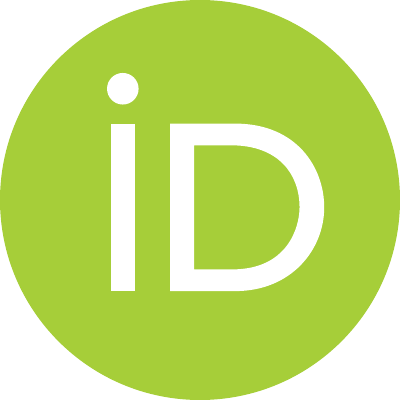}}}
\email{pratik@yukawa.kyoto-u.ac.jp}
\affiliation{Center for Gravitational Physics and Quantum Information, Yukawa Institute for Theoretical Physics,\\ Kyoto University, Kitashirakawa Oiwakecho, Sakyo-ku, Kyoto 606-8502, Japan}
\affiliation{RIKEN Interdisciplinary Theoretical and Mathematical Sciences Program (iTHEMS),
Wako, Saitama 351-0198, Japan}
\author{Tanay Pathak\,\,\href{https://orcid.org/0000-0003-0419-2583}
{\includegraphics[scale=0.05]{orcidid.pdf}}}
\email{pathak.tanay@yukawa.kyoto-u.ac.jp}
\affiliation{Center for Gravitational Physics and Quantum Information, Yukawa Institute for Theoretical Physics,\\ Kyoto University, Kitashirakawa Oiwakecho, Sakyo-ku, Kyoto 606-8502, Japan}
\author{Zhuo-Yu~Xian\,\,\href{https://orcid.org/0000-0002-2245-0059}
{\includegraphics[scale=0.05]{orcidid.pdf}}\,}
\email{zhuo-yu.xian@physik.uni-wuerzburg.de}
\affiliation{Institute for Theoretical Physics and Astrophysics and W{\"u}rzburg-Dresden Cluster of Excellence ct.qmat,\\
Julius-Maximilians-Universit{\"a}t W{\"u}rzburg, Am Hubland, DE-97074 W{\"u}rzburg, Germany}
\affiliation{Department of Physics, Freie Universit{\"a}t Berlin, Arnimallee 14, DE-14195 Berlin, Germany}
\author{Johanna Erdmenger\,\,\href{https://orcid.org/0000-0003-4776-4326}
{\includegraphics[scale=0.05]{orcidid.pdf}}\,}
\email{erdmenger@physik.uni-wuerzburg.de}
\affiliation{Institute for Theoretical Physics and Astrophysics and W{\"u}rzburg-Dresden Cluster of Excellence ct.qmat,\\
Julius-Maximilians-Universit{\"a}t W{\"u}rzburg, Am Hubland, DE-97074 W{\"u}rzburg, Germany}

\begin{abstract}
We propose a tridiagonalization approach for non-Hermitian random matrices and Hamiltonians using singular value decomposition (SVD). This technique leverages the real and non-negative nature of singular values, bypassing the complex eigenvalues typically found in non-Hermitian systems. We analyze the tridiagonal elements, namely the Lanczos coefficients and the associated Krylov (spread) complexity, appropriately defined through the SVD, across several examples, including Ginibre ensembles and the non-Hermitian Sachdev-Ye-Kitaev model. We demonstrate that in chaotic cases, the complexity exhibits a distinct peak due to the repulsion between singular values, a feature absent in integrable cases. Using our approach, we analytically compute the Krylov complexity for two-dimensional non-Hermitian random matrices within a subset of non-Hermitian symmetry classes, including time-reversal, time-reversal$^{\dagger}$, chiral, and sublattice symmetry.

\end{abstract}

~~~~~~~~~~~~YITP-24-132, RIKEN-iTHEMS-Report-24

\maketitle

\section{Introduction}
Although chaos is well understood in classical systems, its definition in the quantum domain remains elusive, often relying on statistical correlations among the eigenvalues of the Hamiltonian $H$ describing the system  \cite{haake1991quantum, PhysRevLett.52.1, Wigner1, Dyson1962a, Oganesyan:2007wpd, Atas2013distribution, Brezin1997, Cotler2017, delCampo:2017bzr}. The eigenvalues $E_i$ are obtained by standard diagonalization $H = W \Lambda W^{\dagger}$, where $W$ is a unitary and $\Lambda = \mathrm{diag}(E_1, \cdots, E_d)$, where $d$ is the dimension of the matrix. Instead of diagonalization, it is often useful to consider \emph{tridiagonalizing} the Hamiltonians, as an efficient alternative. This approach has two clear advantages. First, finding the eigenvalues through tridiagonalization is substantially faster. Second, from a physics perspective, utilizing powerful recursion methods, this approach broadly aims to `reduce practically any problem to one dimension', with a chain model represented by a Hamiltonian where only nearest-neighbor interactions are present \cite{viswanath2008recursion}.  For Hermitian systems, tridiagonalization can be achieved as $H = P \Lambda_h P^{\dagger}$, where $P$ is unitary and $\Lambda_h$ is a tridiagonal matrix whose eigenvalues are identical to those of $H$. The tridiagonal matrix elements, known as the Lanczos coefficients, are 
straightforwardly obtained in the
Krylov space approach \cite{lanczos1950iteration, viswanath2008recursion}. These coefficients describe hopping amplitudes in a chain known as the Krylov chain and are crucial for revealing how the eigenvalue correlations manifest in early- and late-time dynamics of complex quantum systems. Understanding these dynamics for unitary evolution has been at the forefront of active research in recent years \cite{Parker:2018a, Barbon:2019wsy, Dymarsky:2019elm, Jian:2020qpp,   Rabinovici:2020ryf, Cao:2020zls, Dymarsky:2021bjq,  Kar:2021nbm, Caputa:2021sib, Rabinovici:2021qqt,  Balasubramanian:2022tpr,   Hornedal:2022pkc, Bhattacharjee:2022vlt, Bhattacharjee:2022qjw, Balasubramanian:2022dnj, Avdoshkin:2022xuw, Erdmenger:2023wjg,  Craps:2023ivc,  Balasubramanian:2023kwd}; see \cite{Nandy:2024htc} for a comprehensive review.

In certain scenarios, for instance, for a system interacting with a dissipative environment, the Hermiticity condition must be relaxed. Under certain conditions such as the presence of a Markovian environment, such a system is described by an effective non-Hermitian Hamiltonian, leading to entirely new dynamics in Krylov space \cite{Bhattacharya:2022gbz, Liu:2022god, Bhattacharjee:2022lzy, Bhattacharya:2023zqt,  NSSrivatsa:2023pby,
Bhattacharjee:2023uwx, Beetar:2023mfn, Bhattacharya:2023yec, Carolan:2024wov, Bhattacharya:2024hto}. Due to non-Hermiticity, the eigenvalues become complex and are distributed in the complex plane. This results in two-dimensional level statistics \cite{mehta1991random, Hamazaki_2020, ComplexspacingProsen, ProsenDSFF}. In such cases, the underlying physics often becomes obscure due to, for example, the cubic level repulsion irrespective of the universality classes in random matrix theory (RMT) \cite{GHconjecture1, PhysRevE.55.205, Hamazaki_2020}. Additionally, the parametrization in a two-dimensional plane \cite{ComplexspacingProsen}, the use of complex time coordinates in temporal evolution \cite{ProsenDSFF}, and finite-size rescalings make the analysis challenging \cite{PhysRevLett.126.090402}.  To address this, Ref.\,\cite{KawabataSVD23} proposed considering the statistics of singular values of the non-Hermitian Hamiltonian instead of its complex eigenvalues. 

Singular values are real and non-negative, simplifying the corresponding statistics to be one-dimensional. Singular values are derived from the singular value decomposition (SVD) of generic non-Hermitian Hamiltonians $H = U \Sigma V^{\dagger}$ (see Table \ref{tab}), where the unitary matrices $U$ and $V^{\dagger}$ are formed by the orthonormal eigenvectors of $H H^{\dagger}$ and $H^{\dagger} H$, respectively. The elements $\sigma_i$ of the diagonal matrix $\Sigma = \mathrm{diag}(\sigma_1, \cdots, \sigma_d)$ are known as the \emph{singular values} of $H$. Alternatively, they are equal to the eigenvalues of $\sqrt{H^{\dagger} H}$ or $\sqrt{H H^{\dagger}}$, as can be shown by using the Hermitization method \cite{FEINBERG1997579} which involves embedding the Hamiltonian into a space of twice its original dimension. The singular value statistics have proven to be powerful enough \cite{KawabataSVD23} to probe the 38-fold symmetry classification \cite{PhysRevX.9.041015} in non-Hermitian random matrices, extending beyond the Altland-Zirnbauer (AZ) tenfold symmetry classification \cite{AZclassification} in the Hermitian case.  

\begin{table}
\begin{tabular}{ |c|c|c| c| } 
\hline
 & Hamiltonian & Diagonalization & Tridiagonalization \\
\hline
EVD & $H = H^{\dagger}$ & $H = W \Lambda W^{\dagger}$ & $H = P \Lambda_h P^{\dagger}$  \\ 
SVD & $H \neq H^{\dagger}$ & $H = U \Sigma V^{\dagger}$ & $\textcolor{blue}{H = S \,\Sigma_{h} T^{\dagger}}$ \\ 
\hline
\end{tabular}
\caption{The table illustrates the eigenvalue decomposition (EVD) and singular value decomposition (SVD) for Hermitian and non-Hermitian Hamiltonians. A single unitary matrix suffices to diagonalize or tridiagonalize a Hermitian matrix, while two unitary matrices are required for non-Hermitian cases. The expression $H = S \,\Sigma_{h} T^{\dagger}$ is one of the main results of this paper.} \label{tab}
\end{table}

Compared to complex spectral statistics, this simplification through SVD allows Hermitian probes to be straightforwardly extended to non-Hermitian settings by substituting eigenvalues with singular values \cite{KawabataSVD23, ChenuSVD, Hamanaka:2024njv, Nandy:2024wwv, Tekur:2024kyt}. In the Hermitian limit, singular values reduce to the absolute values of the eigenvalues. This raises several intriguing questions: Is it possible to tridiagonalize non-Hermitian Hamiltonians while preserving the singular values? Can we use this framework to distinguish the dynamics in integrable and quantum chaotic systems in the non-Hermitian case?

The structure of the paper is as follows. In Sec.\,\ref{sec:method}, we introduce the Krylov space approach to SVD to compute the Lanczos coefficients and Krylov complexity using the singular value spectrum. In Sec.\,\ref{example}, we explore three different examples, including Ginibre ensembles and the non-Hermitian Sachdev-Ye-Kitaev (SYK) model, and compute the Krylov (spread) complexity in these contexts. We discuss the emergence of the peak and its relation to the repulsion between singular values in the non-Hermitian setting. Section \ref{symmetry} discusses the symmetry classification of non-Hermitian random matrices from the perspective of Krylov (spread) complexity, providing analytic examples in two dimensions and numerical results for the non-Hermitian SYK model. 
We conclude with the summary and a brief outlook in Sec.\,\ref{conc}.

\section{Methods and Formalism} \label{sec:method}

In this paper, we introduce a framework to address the aforementioned questions, with a particular focus on applications to physical systems. The advantage of our method lies in its ability to tridiagonalize via SVD, bypassing the complexities associated with tridiagonlization using complex eigenvalues \cite{Bhattacharya:2023zqt}. The key insight is to reformulate the tridiagonalization problem through eigenvalue decomposition (EVD) of $\sqrt{H^{\dagger} H}$ or $\sqrt{H H^{\dagger}}$. In other words, the Hamiltonian in question can be expressed as $H = S \Sigma_h T^{\dagger}$, where $\Sigma_h$ is a tridiagonal matrix and $S$ and $T$ are unitary matrices. As highlighted in Table \ref{tab}, $\Sigma_h$ and $T$ are determined by tridiagonalizing $\sqrt{H^\dagger H}= T \Sigma_h T^\dagger$, followed by $S=H T \Sigma_h^{-1}$. Such transformation can be readily implemented using Householder reflections \cite{householder} and using the command $\mathtt{HessenbergDecomposition[\sqrt{H^{\dagger}.H}]}$ in Mathematica which has the default choice of initial state as $(1,0,\cdots,0)^{\intercal}$. It is important to note that this decomposition is not unique, as one has the freedom to reshuffle the singular values and choose the appropriate eigenvector phases. Nevertheless, it ensures that the singular values of $H$ and $\Sigma_h$ remain identical, recasting the eigenvalue tridiagonalization to the singular values.

Consequently, the above procedure suggests the time evolution of the state of the form $\ket{\Psi(t)} = e^{-i \sqrt{H^{\dagger} H} t} \ket{\Psi(0)}$ with an initial state $\ket{\Psi(0)}$ at $t = 0$ \cite{ChenuSVD}. Note that although $H$ is non-Hermitian, the combination $\sqrt{H^{\dagger} H}$ is Hermitian; thus the evolution still remains unitary. Nonetheless, the non-Hermiticity of $H$ will affect the time evolution in a complex manner. Since our focus is on the dynamics governed by the singular values, we aim to capture these dynamics using our proposed framework.

Since $\sqrt{H^{\dagger} H}$ is Hermitian, the Lanczos algorithm \cite{viswanath2008recursion, Balasubramanian:2022tpr} can be straightforwardly applied to produce the Krylov basis $\{(\sqrt{H^{\dagger} H})^{n} \ket{K_0}, n \in N \}$, where $\ket{K_0} = \ket{\Psi(0)}$.  Expressing it in the Krylov basis, $\sqrt{H^{\dagger} H} \ket{K_n} = \mathsf{b}_n \ket{K_{n-1}} + \mathsf{a}_n \ket{K_{n}} + \mathsf{b}_{n+1} \ket{K_{n+1}}$, provides a set of Lanczos coefficients $\{\mathsf{a}_n, \mathsf{b}_n\}$. Alternatively, one may generate the same Krylov space by $(\sqrt{H H^{\dagger}})^{n}$, with a different choice of initial state given by $ST^\dagger\ket{\Psi(0)}$. Of particular interest is the initial state being the superposition state $\ket{\Psi(0)} = (\nicefrac{1}{\sqrt{Z_{\beta}}})\sum_i e^{-\beta \sigma_i/2} \ket{v_i}$, where $\{\sigma_i, \ket{v_i}\}$ are the eigenvalues and eigenstates of $\sqrt{H^{\dagger} H}$ and $Z_{\beta} = \sum_i e^{-\beta \sigma_i}$ is the partition function at temperature $\beta^{-1}$.

Given the Lanczos coefficients, we solve the recursion relation $i  \partial_t \uppsi_n (t) = \mathsf{b}_{n} \uppsi_{n-1}(t) + \mathsf{a}_{n}\uppsi_n(t) + \mathsf{b}_{n +1} \uppsi_{n +1}(t)$ for $n \leq d$ to obtain $\uppsi_n(t)$, known as the Krylov space wave functions \cite{Parker:2018a, Balasubramanian:2022dnj}. The recursion, being equivalent to the Schr\"odinger equation,  describes a one-dimensional particle hopping problem in the Krylov chain with the on-site potential and amplitudes given by $\mathsf{a}_n$ and $\mathsf{b}_n$, respectively. The average position of the particle in this chain is given the Krylov (spread) complexity $  K_S(t) = \sum_n n \,|\uppsi_n (t)|^2$ \cite{Parker:2018a, Balasubramanian:2022tpr}. Through several examples, we demonstrate that in chaotic non-Hermitian systems, this quantity exhibits a distinct ramp-peak-slope-plateau pattern, which is analogous to a slope-dip-ramp-plateau structure in the singular form factor \cite{ChenuSVD, Nandy:2024wwv}. The presence of a peak indicates the repulsion between singular values, a feature absent in integrable Hamiltonians.

Building upon the above framework, we further assert that the density of singular values of any non-Hermitian Hamiltonian is related to the mean Lanczos coefficients $\mathsf{a}(x)$ and $\mathsf{b}(x)$ by the form
\begin{align}
    \uprho (\sigma) = \int_{0}^1 dx \,\frac{\Theta(4 \mathsf{b}(x)^2 - (\sigma - \mathsf{a}(x))^2)}{\pi \sqrt{4 \mathsf{b}(x)^2 - (\sigma - \mathsf{a}(x))^2}}\,, \label{den0}
\end{align}
akin to the Hermitian cases \cite{Balasubramanian:2022dnj, Balasubramanian:2023kwd}, which is based on the seminal work by Dumitriu and Edelman \cite{Dumitriu:2002beta}. Notably, such relations are known to exist from the study of orthogonal polynomials \cite{BESSIS1980109}. Here, the mean coefficients and the density are \emph{ensemble-averaged} quantities, and $\Theta(z)$ represents the Heaviside theta function. We assume the existence of a continuous large-$d$ (thermodynamic) limit, where $x = n/d$ is a scaled variable with respect to the matrix dimension $d$. Note that \eqref{den0} is valid in the \emph{bulk} of the Lanczos spectrum, and the appropriate \emph{padding} at the \emph{edges} $n \sim o(1)$ is achieved through the moment method \cite{Balasubramanian:2022dnj, Balasubramanian:2023kwd, Nandy:2024zcd} described later.
\section{Models and examples} \label{example}
We support our claim by providing three prototypical examples: $(i)$ the non-Hermitian real Ginibre ensemble and $(ii)$ a model interpolating between integrable and chaotic regimes, and $(iii)$ the non-Hermitian Sachdev-Ye-Kitaev (SYK) model \cite{PhysRevLett.70.3339, Kittu}. We describe the results for each of the models separately.

\subsection{Model 1: Ginibre Orthogonal Ensemble (GinOE)} The Ginibre ensemble consists of a set of matrices whose elements have zero mean and variance $1/d$, where $d$ is the size of the matrix \cite{Ginibre:1965zz}. We take $H \in \mathrm{GinOE}$, as a member of Ginibre orthogonal ensemble (GinOE), i.e., consisting of real elements. As stated earlier, the SVD provides singular values that are non-negative and real. The density of the singular values is well described by the \emph{quadrant law} \cite{SHEN20011}
\begin{align}
    \uprho (\sigma) = \frac{1}{\pi} \sqrt{4-\sigma^2} \,, ~~~~~ \sigma \in [0,2]\, ,\label{quad}
\end{align}
normalized within the interval. Compare this to the semicircle distribution of the eigenvalues of Hermitian matrices $\uprho_{\mathrm{sc}} (E) = (\nicefrac{1}{2 \pi})\sqrt{4-E^2}$ with $E \in [-2,2]$. We immediately see that \eqref{quad} lacks the parity symmetry $\sigma\rightarrow - \sigma$ which is present in $\uprho_{\mathrm{sc}} (E)$ for $E \rightarrow -E$. This restricts the corresponding histogram at the first quadrant, providing Lanczos coefficients via \eqref{den0}, which exhibits different behavior compared to the Hermitian counterpart.

\begin{figure}[t]
\hspace*{-0.5cm}
\includegraphics[width=0.43\textwidth]{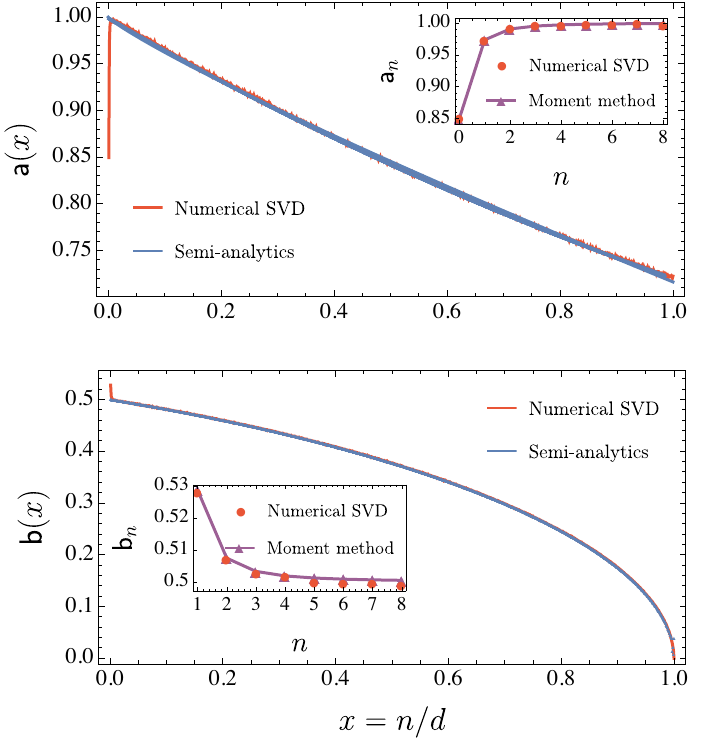}
\caption{Diagonal (top) and off-diagonal (bottom) Lanczos coefficients for the GinOE ($d = 1024$ with $1000$ independent Hamiltonian realizations) computed from the numerical SVD, the fitting function \eqref{bns}, and the semianalytic approach \eqref{den0}. The insets show the agreement between the Lanczos coefficients at edges $n \sim o(1)$, obtained by the moment method.}\label{fig:Lanczos}
\end{figure}

Figure \ref{fig:Lanczos} illustrates the mean value of the diagonal and off-diagonal Lanczos coefficients, initiated with the state $(1,0,\cdots,0)^{\intercal}$. We observe distinct features from the Hermitian counterparts. The diagonal elements $\mathsf{a}_n$ are nonvanishing which stems from the fact that \eqref{quad} is not symmetric in $\sigma \rightarrow - \sigma$. The off-diagonal elements $\mathsf{b}_n$ follow a decreasing pattern and vanish at the end of the Krylov space. The numerical fit suggests the following behavior of the coefficients in the bulk,
\begin{align}
    \mathsf{a}(x) \simeq 1 - p \,x^{q}\,,~~~~
    \mathsf{b}(x) \simeq \frac{1}{2} \sqrt{1-x^{\gamma}}\,, \label{bns}
\end{align}
where $p\, \approx 0.28$, $q \approx 0.88$, and $\gamma \approx 1.2$. To test the validity of \eqref{bns}, we solve the integral equation \eqref{den0} using the given density \eqref{quad}. The results, superimposed in Fig.\,\ref{fig:Lanczos}, show perfect agreement between the diagonal and off-diagonal coefficients in the bulk, supporting \eqref{bns}. However, Eq.\,\eqref{den0} does not capture the edge $n \sim o(1)$ and the tail $d - n \sim o(1)$ of the spectrum \cite{Balasubramanian:2022dnj, Balasubramanian:2023kwd}.

To examine the detailed behavior at the edge $n \sim o(1)$, we apply the moment method \cite{Nandy:2024htc}. The moments of the density of states \eqref{quad} are given by $m_k = \int_0^2 \sigma^k \uprho (\sigma)\, d \sigma = \mathrm{C}_{k/2}$, where $\mathrm{C}_{k/2}$ is the \emph{half-integer} Catalan number. Although Catalan numbers are primarily defined for integer values, the half-integer ones can be defined by expressing the integer Catalan number in terms of the Gamma function and extending it to noninteger values, i.e., $ \mathrm{C}_{k/2} = \frac{2^k \Gamma(\frac{k}{2} + \frac{1}{2})}{\sqrt{\pi} \Gamma(\frac{k}{2} + 2)}$, with $k \in N$. For the semicircle distribution supported on $E \in [-2,2]$, the odd moments vanish and the even moments are given by integer Catalan numbers. However, in our case, due to the presence of odd moments, $\mathsf{a}_n$ will be nonvanishing. The insets in Fig.\,\ref{fig:Lanczos} show the behavior of the edge coefficients computed from the moment method, which are appropriately \emph{matched} or \emph{padded}  \cite{Balasubramanian:2023kwd, Bhattacharjee:2024yxj, Nandy:2024zcd} with the bulk coefficients. 

We briefly remark on the justification of this matching. The moment method, which assumes an infinite-dimensional system, does not yield accurate $n \sim o(d)$ coefficients for finite-dimensional systems. Therefore, there exists a regime of $n \sim o(1)$ in the large-$d$ limit where the Lanczos coefficients computed using these two methods are appropriately {matched}; this matching is referred to as  \emph{padding}. This ensures that the full numerical Lanczos coefficients accurately describe the analytical or semianalytical results. Notably, this padded regime does not scale with the system size $d$ in the large-$d$ limit. Padding at  $n \sim o(1)$ is necessary because the edge coefficients cannot be determined from the integral equation \eqref{den0}. We compare the bulk and edge coefficients with the numerical SVD results (highlighted in red in Fig.\,\ref{fig:Lanczos}) and observe perfect agreement between them.

\subsection{Model 2: A linear combination of integrable and GinUE} Our second example considers a Hamiltonian as a linear combination of the form
\begin{align}
    H = (1-\nu) P + \nu Q\,, \label{haminterpol}
\end{align}
where $P$ is an uncorrelated diagonal matrix with complex entries and $Q \in \mathrm{GinUE}$, i.e., the Ginibre unitary ensemble \cite{Ginibre:1965zz}. Both matrices are non-Hermitian, with elements given by $\langle P_{mn} \rangle = \langle Q_{mn} \rangle = 0$, $\langle |P_{nn}|^2 \rangle = 1$, and $ \langle |Q_{nn}|^2 \rangle = 1/d$. Here $0 \leq \nu \leq 1$ is a parameter that interpolates the Hamiltonian between the regimes governed by $P$ and $Q$ matrices, respectively. We focus on two extreme limits $\nu = 0$ and $\nu = 1$, where $H$ is uncorrelated and GinUE.

The quantity of our interest is the singular-spacing ratio 
$\langle r_{\upsigma} \rangle = \mathrm{mean} (r_{\upsigma,n})$, with $r_{\upsigma,n} = \frac{\mathrm{min}(\lambda_{n}, \lambda_{n+1})}{\mathrm{max}(\lambda_{n}, \lambda_{n+1})}$ \cite{KawabataSVD23}. Here $\lambda_n = \sigma_{n+1} - \sigma_n$ denotes the spacing between the singular values $\{\sigma_i\}$ of the Hamiltonian $H$. As demonstrated in \cite{KawabataSVD23, ChenuSVD, Hamanaka:2024njv, Nandy:2024wwv}, such spacing distributions, including the higher-spacing ratios \cite{Tekur:2024kyt}, generalize the usual level spacing distribution in a much more transparent way compared to the complex spacing ratios \cite{ComplexspacingProsen}. In other words, the underlying chaotic and integrable properties of the system are well described by the level repulsion and clustering between the singular values. Numerically, we find $\langle r_{\upsigma} \rangle \simeq 0.38$ for $\nu = 0$ (integrable), and $\langle r_{\upsigma} \rangle \simeq 0.60$ for $\nu = 1$ (chaotic).

\begin{figure}[]
\hspace*{-0.4cm}
\includegraphics[width=0.45\textwidth]{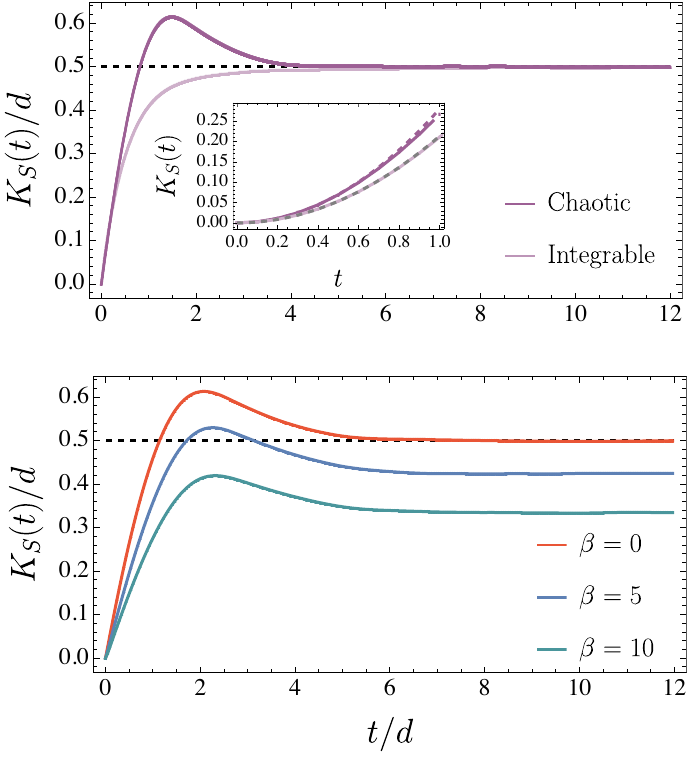}
\caption{Top: Behavior of the Krylov (spread) complexity for the Hamiltonian \eqref{haminterpol} in two regimes: $\nu = 0$ (integrable, light purple) and $\nu = 1$ (chaotic, dark purple). The initial state is the superposition state with $d = 500$ for $200$ independent Hamiltonian realizations. In the chaotic regime, the complexity exhibits a prominent peak, which is absent in the integrable regime. Inset shows the early-time quadratic behavior of the complexity. Bottom: The behavior of the complexity for the non-Hermitian SYK$_4$ model \eqref{nhsykhsparse} with $N = 22$ ($d = 2^{N/2-1}$ is the dimension of the parity-symmetry-resolved block) and $100$ independent Hamiltonian realizations for different inverse temperatures $\beta$, initiated with a superposition state. The saturation gets lower as temperature decreases.}\label{fig:kcompGin}
\end{figure}
Correspondingly, Fig.\,\ref{fig:kcompGin} (top) illustrates the behavior of Krylov (spread) complexity for the Hamiltonian \eqref{haminterpol}, initiated with a uniform superposition of the eigenstates of that Hamiltonian at $\beta=0$, which is analogous to the infinite-temperature thermofield double (TFD) state in the double-copy Hilbert space. In the chaotic regime ($\nu = 1$), the complexity exhibits a distinct ramp-peak-slope-plateau pattern. This peak arises from the level-repulsion between singular values. This behavior is reminiscent of the linear ramp observed in the singular form factor for non-Hermitian chaotic Hamiltonians \cite{ChenuSVD, Nandy:2024wwv}.

We argue that although the peak is a well-known observation in Hermitian chaotic systems \cite{Balasubramanian:2022tpr}, it is highly nontrivial in non-Hermitian scenarios. The peak, expected to arise from eigenvalue repulsion, has no \emph{a priori} reason to manifest as a correlation between complex eigenvalues \cite{ComplexspacingProsen}. Moreover, the dissipative spectral form factor \cite{ProsenDSFF}, computed using such complex eigenvalues, shows a \emph{quadratic} ramp instead of a linear one. Therefore, there is no straightforward way to associate such a peak with the ramp in non-Hermitian cases. Intriguingly, these correlations become transparent when quantities are expressed in terms of singular values, such as the appearance of the \emph{linear} ramp in the singular form factor \cite{ChenuSVD, Nandy:2024wwv}. Thus, the appearance of the peak through our tridiagonalization procedure provides the \emph{simplest} way to demonstrate the existence of nontrivial correlations between the observed peak and the repulsion between singular values and the associated singular form factor in non-Hermitian settings.

In contrast, the peak is absent in the integrable regime ($\nu = 0$), indicating the level clustering between singular values. The inset of Fig.\,\ref{fig:kcompGin} (top) shows the early-time behavior of complexity, which is quadratic and primarily governed by the off-diagonal coefficients. In both integrable and chaotic regimes, this behavior is approximated by $K_S(t) \approx \mathsf{b}_1^2 t^2$. This initial growth can be understood from the Ehrenfest equation in Krylov space \cite{Erdmenger:2023wjg},
\begin{align*}
&\partial_t^2 K_S(t)= \sum_{n=0}2\kc{\mathsf{b}_{n+1}^2-\mathsf{b}_n^2}\abs{\uppsi_n(t)}^2 \\
&+(\mathsf{a}_{n+1}-\mathsf{a}_n)\mathsf{b}_n\kc{\uppsi_n(t)\uppsi_{n+1}(t)^*+\uppsi_{n+1}(t)\uppsi_n(t)^*}\,. 
\end{align*}
Initially, the Krylov wave function $\uppsi_n(t)$ is localized at $n=0$, i.e., $\uppsi_n(0)=\delta_{n0}$, and the above equation reduces to $\partial_t^2 K_S(0)= 2\mathsf{b}_1^2$. The $\mathsf{a}_n$ coefficients hardly have any impact at the early time. The plateau value is independent of the chaotic and integrable nature of the Hamiltonian and depends on the dimension of the Hilbert space. For infinite-temperature TFD state, the plateau is $\bar{K}_S/d = (d-1)/(2d)$, akin to the Hermitian case \cite{Erdmenger:2023wjg}.


\subsection{Model 3: Non-Hermitian SYK model} Our final example is the non-Hermitian SYK$_4$ (i.e., $q = 4$) model, given by the  Hamiltonian \cite{Garcia-Garcia:2021rle, Garcia-Garcia:2022xsh}
\begin{align}
  H =   \sum_{1 \leq a < b < c < d \leq N} (J_{abcd} + i M_{abcd}) \,\psi_{a} \psi_{b} \psi_{c} \psi_{d}\, ,\label{nhsykhsparse}
\end{align}
where $\psi_a$ are the Majorana fermions with $\ke{\psi_a,\psi_b}=\delta_{ab}$ and the Gaussian random variables $J_{abcd}$ and $M_{abcd}$ with zero mean and variance $\langle J_{abcd}^2 \rangle = \langle M_{abcd}^2 \rangle = 6/N^3$, respectively. The couplings $M_{abcd}$ explicitly break the Hermiticity of the Hamiltonian. Lately, the entanglement and spectral properties of this model \cite{Cipolloni:2022fej}, and its PT-symmetric \cite{GarciaGarciaPT1, GarciaGarciaPT2, GarciaGarciaKeldysh} and sparse \cite{Nandy:2024wwv} variants have garnered substantial interest, especially due to its potential relation with holographic correspondence \cite{Maldacena:2016remarks}.

Figure \ref{fig:kcompGin} (bottom) illustrates the behavior of the Krylov (spread) complexity for the non-Hermitian SYK$_4$ model of $N = 22$ with $100$ independent Hamiltonian realizations.  For this $N$ (and $q=4$), the model belongs to the non-Hermitian A symmetry class \cite{Garcia-Garcia:2021rle} (analytic results for the $2$-dimensional case on this class are given later), which can also be determined from the $\langle r_{\upsigma} \rangle$ value and average normalized variance \cite{KawabataSVD23}. We focus on a particular parity-symmetric block, reducing the effective dimension $d = 2^{N/2-1}$. We initiate with different superposition states of inverse temperatures $\beta$. Due to the chaotic nature of the model, as evidenced by the singular value statistics and spacing ratios \cite{Nandy:2024wwv}, the complexity exhibits a peak, justifying our preceding argument and the ramp pattern observed in the singular form factor for this model \cite{Nandy:2024wwv}. The saturation value of the plateau decreases as the temperature lowers. Across all three examples, we demonstrate that our proposed framework is robust enough to distinguish the dynamics of integrable and chaotic systems, consistent with the singular value statistics.

\section{Krylov complexity and symmetry classification} \label{symmetry} 

\begin{table}[]
    \centering
    \begin{tabular}{llll}
    \hline
    Non-Hermitian class & ~~~Hermitization & ~~~~~$\upbeta$ & ~~~~~~\,$\upalpha$ \\
    \hline
    A   & ~~~~~~~~AIII  & ~~~~~2 & ~~~~~~~1 \\
    AI  & ~~~~~~~~BDI   & ~~~~~1 & ~~~~~~~0 \\
    AI$^\dagger$    &  ~~~~~~~~~CI   & ~~~~~1 & ~~~~~~~1 \\
    AII &   ~~~~~~~~\,CII    &  ~~~~~4   & ~~~~~~~3 \\
    AII$^\dagger$ & ~~~~~~~~DIII    & ~~~~~4 & ~~~~~~~1 \\
    AIII &~~~~~~~~~\,A & ~N/A(A) & ~~~N/A(A) \\
    AIII$^\dagger$ &~~~~AIII$\times$AIII & ~N/A(A) & ~~~~~~~1 \\
    D   & ~~~~~~~~DIII  & ~~~~~4 & ~~~~~~~1 \\
    C   & ~~~~~~~~\,CI    & ~~~~~1 & ~~~~~~~1 \\
    \hline
    \end{tabular}
    \caption{Singular-value statistics in classes of non-Hermitian matrices as given in \cite{KawabataSVD23}. Here ``N/A(A)'' means that the singular-value statistics are not characterized by $(\upbeta,\upalpha)$ but given by those in Hermitian matrices in class A. AIII$\times$AIII means the Hermitized matrix is decomposed into two independent Hermitian matrices of class AIII. For instance, the Gaussian orthogonal ensemble (GOE) belongs to the Hermitian AI class, while the Ginibre orthogonal ensemble (GinOE) belongs to the non-Hermitian A class. An example of a non-Hermitian AI$^{\dagger}$ class is GOE + $i$ GOE.}
    \label{tab:nHRMT}
\end{table}

The singular-value statistics of all symmetry classes of non-Hermitian random matrices are investigated in \cite{KawabataSVD23}. The joint probability distribution of singular values is given by \cite{KawabataSVD23} 
\begin{align}
    \uprho(\ke{\sigma_i}) = C_N \prod_{i=1}^d \sigma_i^\upalpha \prod_{i<j}^d \abs{\sigma_i-\sigma_j}^\upbeta e^{-\frac{\upbeta d}{4}\sum_{i=1}^d \sigma_i^2}\,. \label{eq:SingularDistribution} 
\end{align}
where $C_N$ is a normalization constant and we rescale the singular values so that the corresponding spectrum density follows the quadrant law \eqref{quad}. Here $\upbeta$ is Dyson's index (not to be confused with the inverse temperature $\beta$) and $\upalpha$ is known as the hard-edge exponent \cite{Sun:2019yqp, Xiao:2024lgi}, characterizing the distribution of the minimal singular value \cite{KawabataSVD23}. The level-spacing distribution of the singular values $\uprho(\lambda)$ weakly depends on $\upalpha$ and strongly depends on $\upbeta$. Using the Hermitization method, the spacing distribution of singular values for non-Hermitian matrices can be mapped \cite{KawabataSVD23} to the tenfold Altland–Zirnbauer (AZ) classes \cite{AZclassification}. For convenience, we list some of these classes and the Hermitized mapping in Table \ref{tab:nHRMT} \cite{KawabataSVD23}. The mapping implies the singular value statistics of the non-Hermitian classes are identical to the singular value statistics of the corresponding Hermitized matrices. For such Hermitization, which falls under Bogoliubov-de-Gennes (BdG) and chiral classes in the AZ classification, the singular value spacing distribution of the corresponding matrices is approximated by the standard Wigner-Dyson distribution \cite{haake1991quantum} \begin{equation}
    \uprho_{\mathrm{WD}}(\lambda)=\frac{2 z_{\upbeta} e^{-(\lambda z_{\upbeta})^2} (\lambda z_{\upbeta})^{\upbeta }}{\Gamma \big(\frac{1+\upbeta}{2}\big)}\,,~~ 
    z_{\upbeta} =\frac{\Gamma \big(1+ \frac{\upbeta }{2}\big)}{\Gamma \big(\frac{1+\upbeta}{2}\big)}\,.
    \label{eq:WD}
\end{equation}
Here we rescaled both the spacing $\lambda$ and the distribution such that $\int_0^\infty \uprho_{\mathrm{WD}}(\lambda) d\lambda = \int_0^\infty \lambda \uprho_{\mathrm{WD}}(\lambda) d\lambda=1$. 

\begin{figure}
    \centering
\includegraphics[width=0.9\linewidth]{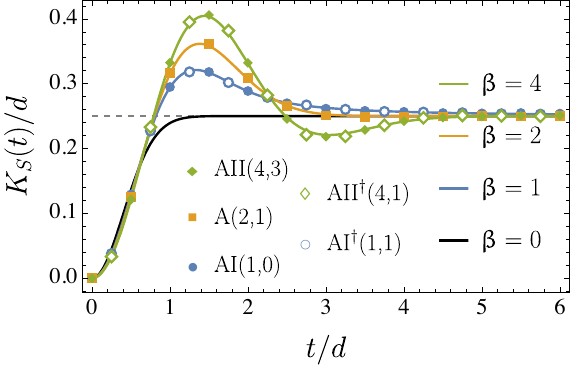}
    \caption{Krylov (spread) complexity for $d=2$ non-Hermitian random matrices in several classes of the 38-fold symmetry classes \cite{PhysRevX.9.041015}. For example, label A\,(2,1) represents the non-Hermitian A symmetry class, whose singular values distribution is parametrized by indices $(\upbeta,\upalpha)=(2,1)$ according to \cite{KawabataSVD23}. The solid curves represent the analytical result \eqref{eq:SKC2DA} with Dyson's index $\upbeta=0,1,2,4$. Here $\upbeta = 0$ refers to the integrable Poisson case \cite{PhysRevLett.122.180601} with $K_S(t) = \nicefrac{1}{2}(1-e^{-\pi t^2/4})$. The GinUE, GinOE, and GinSE (S = symplectic) belong to non-Hermitian symmetry classes A, AI, and AII, respectively. The non-Hermitian SYK$_4$ for $N = 22$ also belongs to A class \cite{Garcia-Garcia:2021rle}.}
    \label{fig:SKC2D}
\end{figure}

To support our numerical results, in particular the existence of the peak in chaotic cases, we turn to analytical examples and provide an exact expression for the Krylov (spread) complexity in $2$-dimensional non-Hermitian random matrices. Suppose $H$ is a $2\times2$ non-Hermitian matrix whose singular values are given by $(\sigma_0,\sigma_1)$. They are also the eigenvalues of $\sqrt{H^{\dagger}H}$. Take the initial state as the superposition state $\ket{\Psi(0)}=(\ket0+\ket1)/\sqrt2$ as the initial state, the Lanczos coefficients are straightforwardly computed: $\mathsf{a}_0= \mathsf{a}_1=\sigma,\, \mathsf{b}_1=\lambda/2$, where $\sigma=(\sigma_0+\sigma_1)/2,\,\lambda=\abs{\sigma_0-\sigma_1}$ denotes the level-spacing between the singular values. The Krylov wave functions are given by $\uppsi_0(t) = e^{-i \sigma  t} \cos (\lambda  t/2)$ and $\uppsi_1(t) = -i e^{-i \sigma  t} \sin (\lambda t/2)$. The complexity is $K_S(t)=\sin^2(\lambda t/2)$, which is periodic and only depends on the level spacing $\lambda$ \cite{Caputa:2024vrn}. Hence, the ensemble average of the Krylov (spread) complexity of $2\times2$ non-Hermitian random matrices, whose Hermitization belongs to the BdG and chiral classes, is described by Dyson's index $\upbeta$ only.
The average Krylov complexity is given by $K_S(t) =\int_0^\infty \sin^2(\lambda t/2) \uprho_\text{WD}(\lambda) \,d\lambda$, resulting in
\begin{align}
    K_S(t)
    =\frac{1}{2}\kc{1- \, _1F_1\left(\frac{1+\upbeta}{2};\frac{1}{2};-\frac{t^2}{4 z_{\upbeta}^2}\right)}\,, \label{eq:SKC2DA}
\end{align}
where $_1F_1$ is the Kummer confluent hypergeometric function. It is to be understood that complexity is evaluated using the ensemble average. We consider non-Hermitian matrices whose Hermitization maps to one of the tenfold AZ classes. Using our tridiagonalization method, with the superposition state as the initial state, we numerically compute the complexity for the five non-Hermitian symmetry classes from Table \ref{tab:nHRMT} [A, AI, AII, AI$^{\dagger}$ and AII$^{\dagger}$ with respective $(\upbeta, \upalpha)$], which is depicted in Fig.\,\ref{fig:SKC2D}. The numerical results show perfect agreement with the analytic result \eqref{eq:SKC2DA} for these symmetry classes considered. The same will be expected to hold for all $38$-fold symmetry classes. We notice that the impact of $\upalpha$ is negligible. The plateau value is $\bar{K}_S/d = (d-1)/(2d) = 0.25$, identical for all classes, where the overline indicates the late-time value. 

\begin{figure}[t]
    \centering
    \includegraphics[width=1\linewidth]{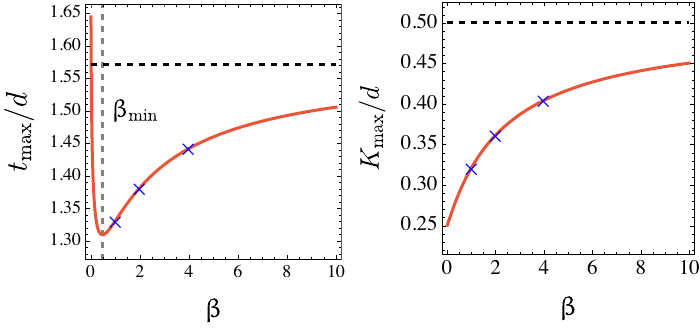}
    \caption{Left: Timescale of the maximal value (left) and the peak value (right) of the complexity \eqref{eq:SKC2DA} for different Dyson's index $\upbeta$. Here $\upbeta$ is treated as a continuous parameter, extending to noninteger values \cite{Dumitriu:2002beta, PhysRevLett.122.180601}, including $\upbeta = 1,2,4$ (these specific points are marked with blue crosses). The black dashed line indicates the corresponding values at $\upbeta \rightarrow \infty$.}
    \label{fig:kryana}
\end{figure}

We briefly explore the properties of the analytic result \eqref{eq:SKC2DA}. Although the expression is derived and numerically verified for the standard ensembles with $\upbeta = 1, 2, 4$, one can consider $\upbeta$ to be a continuous variable with $\upbeta \in [0, \infty)$ \cite{PhysRevLett.122.180601}, inspired by the seminal work of Dumitriu and Edelman \cite{Dumitriu:2002beta} in Gaussian $\upbeta$ ensembles. We observe that as $\upbeta$ increases, the complexity exhibits higher peak values. Ensembles with higher-level correlations show more pronounced peaks in complexity. It is reasonable to question whether this peak continues to rise as $\upbeta$ increases further. Conversely, this also suggests that there exists a range of $\upbeta$ below which complexity does not exhibit any peak.

To investigate this, we extend \eqref{eq:SKC2DA} to general $\upbeta \in [0, \infty)$ and numerically determine the maximal value and the timescale at which it is reached (Fig.\,\ref{fig:kryana}). We find that for $\upbeta \geq \upbeta_{\mathrm{min}}$ with $ \upbeta_{\mathrm{min}} \simeq 0.48$, as indicated by the dashed line in Fig.\,\ref{fig:kryana} (left), complexity indeed exhibits a peak. Since the peak only exists for $\upbeta \geq \upbeta_{\mathrm{min}}$, Fig.\,\ref{fig:kryana} (left) is only relevant in this regime. However, the peak value and the associated timescale do not grow indefinitely; they saturate at $K_{\mathrm{max}}/d \simeq 0.5$ and $t_{\mathrm{max}} \simeq 1.57$. The significance of this maximal time scale is unclear at this moment, but likely to represent the peak value for a completely rigid spectrum. In Hermitian settings, such continuous $\upbeta$ ensembles with $\upbeta \in [0, 1]$ can be mapped \cite{PhysRevLett.122.180601} to physical Hamiltonians exhibiting many-body localized (MBL) transitions \cite{MBLpalHuse}. Hence, we expect the transition of $\upbeta$ might hold significance in analogous MBL transitions in non-Hermitian settings \cite{KawabatanonHMBL}.

\begin{figure}
    \centering
\includegraphics[width=0.87\linewidth]{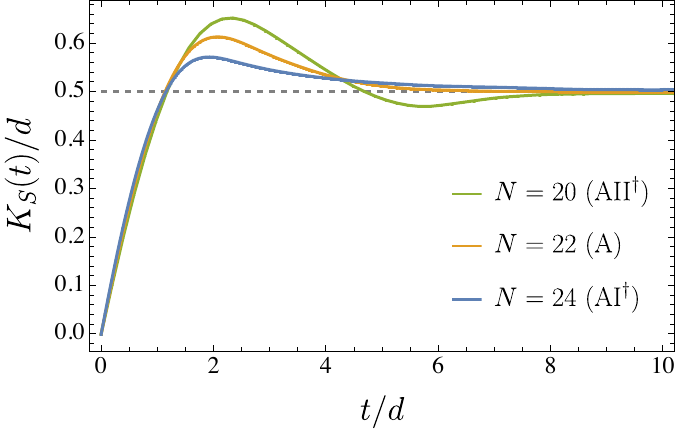}
    \caption{The evolution of Krylov (spread) complexity for the non-Hermitian SYK$_4$ model is shown for various values of $N$. Here, $d = 2^{N/2-1}$ represents the dimension of the parity-symmetry-resolved block. The values of $N$ correspond to different symmetry classes: $\mathrm{AII}^{\dagger}$ ($N = 20$), $\mathrm{A}$ ($N = 22$) and  $\mathrm{AI}^{\dagger}$ ($N = 24$) \cite{Garcia-Garcia:2021rle}. The complexity is calculated using $100$ Hamiltonian realizations for each $N$.}
    \label{fig:SKCkrySYK}
\end{figure}

The non-Hermitian SYK model for large dimensions, which belongs to the corresponding symmetry classes \cite{PhysRevB.95.115150, Garcia-Garcia:2021rle} validates the behavior of the analytic results shown in Fig.\,\ref{fig:SKC2D}. The symmetry classes for the non-Hermitian SYK model are determined by the total number of fermions ($N\, \mathrm{mod}~ 8$) \cite{PhysRevB.95.115150} and the total number of interacting fermions ($q \,\mathrm{mod}~ 4$) \cite{Garcia-Garcia:2021rle}. In our case, with $q=4$, the symmetry classes are as follows: $N = 20 \,(\mathrm{AII}^{\dagger})$, $N = 22 \,(\mathrm{A})$, and $N = 24 \,(\mathrm{AI}^{\dagger})$ (see Table II in \cite{Garcia-Garcia:2021rle}). The results for the SYK model are depicted in Fig.\,\ref{fig:SKCkrySYK}, showing a similar trend to the analytic results for $2\times2$ random matrices in the same symmetry class. The complexity for class $\mathrm{AII}^{\dagger}$ exhibits the highest peak (with a small dip), followed by $\mathrm{A}$ and $\mathrm{AI}^{\dagger}$. The saturation is identical for all three classes, providing a compelling validation of our analytic results for large-$d$ systems.

A further advantage of our approach lies in the following. For non-Hermitian random matrices in all Ginibre ensembles (GinOE, GinUE, and GinSE), the eigenvalues are distributed in the complex plane and exhibit cubic repulsion \cite{GHconjecture1, PhysRevE.55.205}, with separation computed through the Euclidean distance in the complex plane. Their angular distribution must be properly considered to compute the spacing ratios \cite{ComplexspacingProsen}. Consequently, defining the $2 \times 2$ matrix through the complex eigenvalues requires four distinct parameters, making it unclear how eigenvalue separation can be used \cite{GHSconjecture1, GHconjecture1, AkemannProsen} to obtain analytic results for complexity for different symmetry classes. However, in our SVD tridiagonalization approach the analytic expressions are readily obtained, which underscores the usefulness of our approach.

\section{Conclusion and outlook} \label{conc}
By applying Krylov space methods to SVD, we provide an approach for tridiagonalizing non-Hermitian Hamiltonians. This approach preserves the singular values of the original Hamiltonian, similar to how eigenvalues remain unchanged in the Hermitian case. Aligning with the singular value statistics \cite{KawabataSVD23} in non-Hermitian systems, our proposed definition of Krylov (spread) complexity through SVD successfully distinguishes between integrable and chaotic systems, as evidenced by a distinct peak in the latter case. This peak arises due to the level repulsion of singular values, which is related to the linear ramp observed in the singular form factor \cite{ChenuSVD, Nandy:2024wwv}, akin to the spectral form factor in Hermitian cases \cite{Balasubramanian:2022tpr, Erdmenger:2023wjg}. Notably, for two dimensions, we present both numerical and analytical results for non-Hermitian random matrices, marking the observation of a distinct peak for non-Hermitian chaotic systems in different symmetry classes. This peak is clearly explained through the SVD approach and cannot be easily accounted for by correlations between complex eigenvalues. Our Krylov space approach through SVD in various symmetry classes thus provides a clearer picture of how static quantities like level correlations leave an imprint on dynamic quantities such as Krylov complexity in non-Hermitian systems. A brief discussion of the Hermitized version of complexity is presented in Appendix \ref{hermitizedKC}.

Our formalism offers a framework for investigating such systems exhibiting complex phase structures. A prototypical example is the non-Hermitian Rosenzweig-Porter model \cite{nonHRP}, which, unlike its Hermitian counterpart \cite{PhysRev.120.1698}, obstructs the fractal phase due to complex diagonal entries. Our approach, particularly a variation of model 2  defined in \eqref{haminterpol} (with $\nu$ scaling with the system size), may provide insights into whether the singular values and the localization of singular vectors, as well as their interplay with the Krylov basis, can elucidate such phase structures \cite{Bhattacharjee:2024yxj}. In these cases, employing multiseed states \cite{Craps:2024suj} as initial states can be fruitful. A potential way to realize the explicit dependence on the initial state is to consider quantum scars in the non-Hermitian PXP model \cite{nonHscar} and follow our construction, analogous to the Hermitian cases \cite{Bhattacharjee:2022qjw}.

A promising direction is to apply the SVD technique to study operator growth in open quantum systems. In the Markovian approximation, operator growth can be formulated using the Lindbladian approach \cite{Lindblad1976, Gorini}. The Lindbladian is non-Hermitian on the Krylov basis \cite{Bhattacharya:2022gbz, Liu:2022god, Bhattacharjee:2022lzy, Bhattacharya:2023zqt, NSSrivatsa:2023pby, Bhattacharjee:2023uwx} for the double Hilbert space, with appropriate jump operators encoding the dissipative strength of the system and the environment. For chaotic systems, the traceless part of the nonintegrable Lindbladian exhibits universal singular value statistics of non-Hermitian random matrices and is classified according to the respective symmetry classes \cite{KawabataSVD23}. 
It is therefore intriguing to explore how the non-unitary Lindbladian dynamics of operator growth can be realized within this framework. In particular, a key question is to better understand the appearance of a logarithmic timescale and the universality of the observed plateau behavior of Krylov complexity in open quantum systems \cite{Bhattacharjee:2022lzy, Bhattacharjee:2023uwx}. These questions will be addressed in future work.

Finally, for the symmetry classification of non-Hermitian random matrices,  we anticipate, based on our results in Fig.\,\ref{fig:SKC2D} -Fig.\,\ref{fig:SKCkrySYK}, that our approach will open up a new avenue for analyzing large-$d$ systems and the universality of the $38$-fold symmetry classes \cite{PhysRevX.9.041015} from a complexity perspective. The different behavior of complexity for these symmetry classes may be a key to understanding the dynamics of open quantum systems and the nature of dissipative quantum chaos \cite{Kawabata:2022cpr, KawabataSVD23}.


\section{Acknowledgments}  We thank Kohei Kawabata, Tadashi Takayanagi, and Masaki Tezuka for valuable discussions. P.N., T.P., and Z.Y.X.~thank the long-term workshop YITP-T-23-01 held at YITP, Kyoto University, where the work was initiated. P.N. is grateful for the hospitality of Kavli IPMU, University of Tokyo, during the final stages of this work, and ICTS, Bengaluru, for the workshop ``Quantum Information, Quantum Field Theory, and Gravity'' (ICTS/qftg2024/08), where part of this work was presented. The work of P.N. is supported by the Japan Society for the Promotion of Science (JSPS) Grant-in-Aid for Transformative Research Areas (A) “Extreme Universe” No. JP21H05190. The Yukawa Research Fellowship of T.P. is supported by the Yukawa Memorial Foundation and JST CREST (Grant No.\,JPMJCR19T2). Z.Y.X. and J.E.~are supported by Germany’s Excellence Strategy through the Würzburg-Dresden Cluster of Excellence on Complexity and Topology in Quantum Matter-ct.qmat (EXC 2147, Project ID 390858490), and by the Deutsche Forschungsgemeinschaft (DFG) through the Collaborative Research center ``ToCoTronics'', Project ID 258499086—SFB 1170.  Z.Y.X. also acknowledges support from the National Natural Science Foundation of China under Grant No.\,12075298.

\appendix\label{appendix}

\section{Hermitized Krylov complexity} \label{hermitizedKC}
In the main text, we have examined the evolution generated by the Hamiltonian $\sqrt{H^{\dagger} H}$, which shares the same dimension, $d$, as the original Hamiltonian $H$. However, it is also possible to Hermitize the Hamiltonian in $2$ dimensions and study the resulting evolution. Given a non-Hermitian matrix $H$, its Hermitized version is
\begin{equation}
    \tilde H=\begin{pmatrix}
    0 & H \\
    H^\dagger & 0 \\
    \end{pmatrix}\,. \label{Herc}
\end{equation}
The eigenvalues of $\tilde{H}$ are given by the set $\ke{E_i^s=s\sigma_i|i=1,2,\cdots,d,\, s=+,-}$, where ${\sigma_i}$ represents the singular values of $H$, whose distribution follows \eqref{eq:SingularDistribution}. Thus, in the large-$d$ limit, the spectral density of $\tilde H$ follows the semicircle law
\begin{equation}
    \uprho_\text{sc}(E)=\frac1{2\pi}\sqrt{4-E^2},\quad E\in [-2,2].
\end{equation}
The eigenvector associated with $E_i^\pm$ is $\ket{E_i^\pm}=(\ket{u_i},\pm\ket{v_i})$ where $H^\dagger \ket{u_i}=\sigma_i\ket{v_i}$ and $H\ket{v_i}=\sigma_i\ket{u_i}$.


We now consider the Krylov (spread) complexity associated with the Hermitized version of the Hamiltonian, which governs the evolution $e^{-i\tilde Ht}$. We refer to this as Hermitized Krylov (spread) complexity. We discuss several choices for the initial state below.

When the initial state $\ket{\Psi(0)}$ is taken in the subspace $\H_+=\text{span}\ke{\ket{E_i^+}}$, the evolution $e^{-i\tilde Ht}\ket{\Psi(0)}$ is controlled by the positive eigenvalues $\ke{E_i^+=\sigma_i}$. Then, the Hermitized Krylov complexity governed by $\tilde H$ reduces to the singular Krylov complexity governed by $\sqrt{H^\dagger H}$ or $\sqrt{HH^\dagger}$ in the main text. More precisely, their moments are identical, namely, $\bra{\Psi_+}\tilde H^n\ket{\Psi_+}=\bra{\Psi_v}(\sqrt{HH^\dagger})^n\ket{\Psi_v}=\bra{\Psi_u}(\sqrt{H^\dagger H})^n\ket{\Psi_u}$, where $\ket{\Psi_+}=\sum_i c_i\ket{E_i^+},\, \ket{\Psi_v}=\sum_i c_i\ket{v_i}$ and $\ket{\Psi_u}=\sum_i c_i\ket{u_i}$. Similarly, when the initial state is taken in the subspace $\H_-=\text{span}\ke{\ket{E_i^-}}$, the Hermitized Krylov complexity governed by $\tilde H$ reduces to the singular Krylov complexity governed by $-\sqrt{H^\dagger H}$ or $-\sqrt{HH^\dagger}$.

Another case is when the chosen initial state does not belong to either $\H_+$ or $\H_-$. For example, the initial state may be chosen as a superposition of all eigenstates: $|\Psi(0) \rangle =(\nicefrac{1}{\sqrt{2d}}) \sum_{i}\sum_{s=\pm}\ket{E_i^s}
=(\nicefrac{1}{\sqrt{d}})\sum_i (\ket{u_i},0)$, so that all the eigenvalues are involved in the evolution. Since the eigenvalues $\ke{E_i^s}$ of $\tilde H$ are symmetric under the reflection with respect to the origin, i.e., $E_i^s\to -E_i^s$, we have $\mathsf{a}_n=0$. From the semicircle law of the bulk spectral density, we have
\begin{equation}\label{eq:bnGaussian1}
    \mathsf{b}_n=\sqrt{1-\frac{n}{2d}}\,,
\end{equation}
in the ensemble average in the large-$d$ limit. 

The simplest example for the Hermitized Krylov (spread) complexity occurs at $d=1$, which allows us to obtain an analytical expression. In this case, the Hermitized matrix $\tilde H$ is a $2\times 2$ matrix with eigenvalues $\ke{-\sigma,\sigma}$ where the singular values $\sigma$ follow the distribution $\uprho(\sigma)\propto \sigma^{\upalpha} e^{-(2z_{\upalpha}\sigma)^2}$ in Eq.\,\eqref{eq:SingularDistribution}, where $z_\upalpha=\Gamma\kc{1+\upalpha/2}/\Gamma\kc{(\upalpha+1)/2}$ so that the average level spacing is unity. Initializing with $\ket{\Psi(0)}=(\ket{u_1},0)$, the Lanczos coefficients and the complexity for each realization of $\tilde H$ are simply given by $\mathsf{a}_0= \mathsf{a}_1=0,\, \mathsf{b}_1=\sigma$ \cite{Caputa:2024vrn}. The Krylov wave functions are $\uppsi_0(t) = \cos(\sigma t)$ and $\uppsi_1(t) = - i \sin(\sigma t)$ with the initial condition $\uppsi_0(0) = 1$. The Krylov (spread) complexity is $K_H(t)=\sin^2(\sigma t)$. Therefore, the ensemble average of Hermitized Krylov (spread) complexity at $d=1$ is
\begin{align}
    K_H(t)
    &= \frac{1}{2} \left(1-\, _1F_1\left(\frac{\upalpha+1}2;\frac{1}{2};-\frac{t^2}{4 z_{\upalpha}^2}\right)\right)\,. \label{kcompan}
\end{align}
$K_H(t)$ has the same form as $K_S(t)$ in \eqref{eq:SKC2DA} except that $\upbeta$ is replaced by $\upalpha$. The reason is that, for the $2\times2$ Hermitization $\tilde H$, the spacing is given by $\lambda=2\sigma$. So, the level repulsion arises from the repulsion between $\sigma$ and the origin, which is characterized by $\upalpha$ rather than $\upbeta$. As we can see, the peak in \eqref{kcompan} disappears when $\upalpha=0$.

For $d \geq 2$, the behavior should qualitatively differ from the case of $d = 1$, as Dyson's index $\upbeta$ plays an increasingly significant role in level repulsion with increasing $d$. At large $d$, the level-spacing distribution depends solely on $\upbeta$. Since the semicircle law is restored after Hermitization, the Hermitized Krylov complexity converges to the results in Hermitian RMT \cite{Balasubramanian:2022tpr, Balasubramanian:2022dnj, Erdmenger:2023wjg}. Our numerical observations indicate that the Hermitized Krylov complexity is sensitive to $\upbeta$ but not to $\upalpha$ at large $d$.

\bibliography{references}

\end{document}